\documentclass[journal,twoside]{IEEEtran}

\usepackage{color}
\usepackage{amssymb}
\usepackage{rotating}
\usepackage[T1]{fontenc}
\usepackage{lscape}
\usepackage{dblfloatfix}
\usepackage{xcolor}
\usepackage{subcaption}
\usepackage[ruled,vlined]{algorithm2e}
\usepackage[autostyle]{csquotes}
\usepackage{tikz}
\usepackage{acro}
\usepackage{siunitx}
\usepackage{enumitem}
\usepackage{multirow}
\usepackage{cite}
\usepackage{hyperref}
\usepackage{amsmath,amssymb,amsfonts}
\usepackage{tikz}
\usepackage{float}
\usepackage{algorithmic}
\usepackage{graphicx}
\usepackage{textcomp}
\usepackage{xcolor}
\usepackage{url}

\usepackage{siunitx}
\usepackage{lineno,hyperref}
\usepackage{breakurl}
\usepackage{todonotes}

\usepackage[utf8]{inputenc}
\usepackage[T1]{fontenc}
\usepackage{colortbl}
\usepackage{pifont}
\usepackage{rotating}
\usepackage{lineno}

\usepackage{comment}
\def\BibTeX{{\rm B\kern-.05em{\sc i\kern-.025em b}\kern-.08em
    T\kern-.1667em\lower.7ex\hbox{E}\kern-.125emX}}

\ifCLASSINFOpdf
\else
 
\fi

\usepackage[a4paper, total={170mm,257mm}]{geometry}

\begin{document}

%


\title{Filter Bubbles in Recommender Systems: Fact or Fallacy - A Systematic Review}


\author{\IEEEauthorblockN{Qazi Mohammad Areeb\IEEEauthorrefmark{1},
Mohammad Nadeem\IEEEauthorrefmark{2}, 
Shahab Saquib Sohail\IEEEauthorrefmark{3},
Raza Imam\IEEEauthorrefmark{1},
Faiyaz Doctor\IEEEauthorrefmark{45}, 
Yassine Himeur\IEEEauthorrefmark{5},
Amir Hussain\IEEEauthorrefmark{6}, and 
Abbes Amira\IEEEauthorrefmark{7}\IEEEauthorrefmark{8}
}\\
\IEEEauthorblockA{\IEEEauthorrefmark{1}
Mohamed bin Zayed University of Artificial Intelligence, computer vision MBZUAI Abu Dhab Masdar City, Abu Dhabi}\\
\IEEEauthorblockA{\IEEEauthorrefmark{2}Department of Computer Science, Aligarh Muslim University, Aligarh, 202002, India}\\
\IEEEauthorblockA{\IEEEauthorrefmark{3}Department of Computer Science and Engineering, Jamia Hamdard University, New Delhi, 110062, India}\\
\IEEEauthorblockA{\IEEEauthorrefmark{5}School of Computer Science and Electronic Engineering, University of Essex, Wivenhoe Park, Colchester CO4 3SQ, United Kingdom}\\
\IEEEauthorblockA{\IEEEauthorrefmark{4}Edinburgh Napier University, United Kingdom}\\
\IEEEauthorblockA{\IEEEauthorrefmark{5}College of Engineering and Information Technology, University of Dubai, Dubai, UAE}\\
\IEEEauthorblockA{\IEEEauthorrefmark{6}Edinburgh Napier University, United Kingdom}\\
\IEEEauthorblockA{\IEEEauthorrefmark{7}Department of Computer Science, University of Sharjah, Sharjah, United Arab Emirates}\\ 
\IEEEauthorblockA{\IEEEauthorrefmark{8}Institute of Artificial Intelligence, De Montfort University, Leicester, United Kingdom}

}


\maketitle

\begin{abstract}
A filter bubble refers to the phenomenon where Internet customization effectively isolates individuals from diverse opinions or materials, resulting in their exposure to only a select set of content. This can lead to the reinforcement of existing attitudes, beliefs, or conditions. In this study, our primary focus is to investigate the impact of filter bubbles in recommender systems. This pioneering research aims to uncover the reasons behind this problem, explore potential solutions, and propose an integrated tool to help users avoid filter bubbles in recommender systems.
To achieve this objective, we conduct a systematic literature review on the topic of filter bubbles in recommender systems. The reviewed articles are carefully analyzed and classified, providing valuable insights that inform the development of an integrated approach. Notably, our review reveals evidence of filter bubbles in recommendation systems, highlighting several biases that contribute to their existence. Moreover, we propose mechanisms to mitigate the impact of filter bubbles and demonstrate that incorporating diversity into recommendations can potentially help alleviate this issue.
The findings of this timely review will serve as a benchmark for researchers working in interdisciplinary fields such as privacy, artificial intelligence ethics, and recommendation systems. Furthermore, it will open new avenues for future research in related domains, prompting further exploration and advancement in this critical area.
\end{abstract}


\IEEEpeerreviewmaketitle


{\color{black}\section{Introduction}
{\color{black}
The proliferation of the Internet has resulted in an overwhelming abundance of information, necessitating the development of systems that can curate and present tailored options from the vast array of available resources \cite{sayed2021intelligent,atalla2023intelligent}. Recommender Systems (RSs) have emerged as a prominent research area, rapidly advancing in their ability to provide users with personalized recommendations for items of interest \cite{himeur2021survey, varlamis2022smart}. However, as the field of recommendation systems progresses, several critical issues have been identified in the literature \cite{dokoupil2022long, sayed2021intelligent}.
Two widely discussed problems in Recommender System Research (RSR) are the "cold start" issue, which pertains to making recommendations for new or sparse users or items \cite{tahmasebi2021hybrid}, and the sparsity problem caused by the lack of available data for certain users or items \cite{ali2022citation}. Furthermore, scalability \cite{wu2022prediction,sardianos2020rehab} and recency time \cite{wu2022news} have been addressed as additional challenges in RSs. In recent years, privacy concerns have also garnered significant attention due to the susceptibility of RSs to security breaches and privacy threats \cite{arif2021towards,himeur2022blockchain}. The emergence of new tools and techniques has introduced novel privacy considerations for RSs, with biased and fair RSs becoming prominent topics in the privacy domain \cite{boratto2019effect, protasiewicz2016recommender}.
Recommender systems exhibit algorithmic biases that can significantly impact their recommendation outputs, potentially leading to issues such as preference manipulation, threat intelligence, and privacy breaches for users \cite{himeur2022latest}. These biases can arise from various aspects and causes within RSs. For instance, favoring frequently purchased items over more relevant ones can lead to popularity bias \cite{ashraf2023private}. Additionally, position bias, exposure bias, selection bias, demographic bias, and anchoring biases may exist in RSs \cite{chen2020bias}. However, the phenomenon of filter bubbles has not been extensively explored in the context of RSR \cite{gao2022mitigating}.

{\color{black}Olshannikova et al. \cite{olshannikova2022utilizing} propose a social diversification strategy for recommending relevant individuals on platforms like Twitter. Their approach leverages dormant ties, mentions of mentions, and community members within a user's network to offer diverse recommendations and facilitate new social connections.
In a study by Alam et al. \cite{alam2022towards}, biases in news recommender systems are examined using stance and sentiment analysis. By conducting an experiment on a German news corpus focused on migration, the study reveals that these recommender systems tend to recommend articles with negative sentiments and stances against refugees and migration. This reinforces user biases and leads to a reduction in news diversity.
Cai et al. \cite{cai2023causal} address issues like echo chambers and filter bubbles caused by recommender systems by concentrating on estimating the effects of recommending specific items on user preferences. They propose a method based on causal graphs that mitigates confounding bias without requiring costly randomized control trials. Experimental results on real-world datasets validate the effectiveness and efficiency of their approach.
Hildebrandt \cite{hildebrandt2022issue} explores the implications of recommender systems prioritizing sales and ad revenue, which can result in feedback loops, filter bubbles, and echo chambers. The article discusses the economic incentives that influence design decisions and examines proposed EU regulations that aim to address these issues by imposing constraints on targeting and requiring responsible design and deployment of recommender systems.}
\begin{figure*}[ht]
\centering
\includegraphics[width=0.8\linewidth]{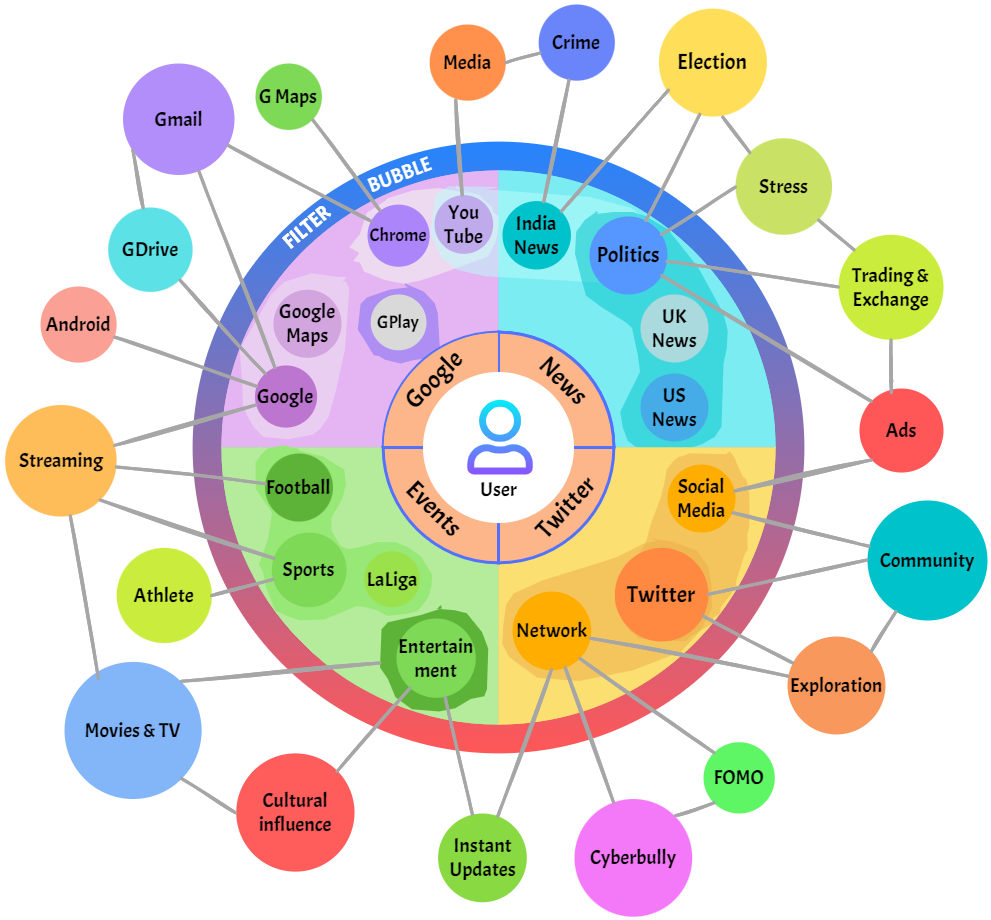}
\caption{Filter bubble}
\label{fig:fb}
\end{figure*}

The investigation of filter bubbles in Recommender Systems (RSs) is a burgeoning area of research that has recently garnered considerable attention, especially in the context of social networks \cite{spohr2017fake}. Initially, there was disagreement regarding the significance of filter bubbles as a problem worthy of attention. However, subsequent discussions in the referenced paper \cite{spohr2017fake} indicate that the majority of practitioners now recognize the importance of addressing this issue. Consequently, there is a consensus that further research is needed to identify effective solutions. Figure \ref{fig:fb} provides a visual representation of a filter bubble.
Given the increasing interest in studying filter bubbles and their impact on recommendation systems, it becomes crucial to conduct a comprehensive Systematic Literature Review (SLR) of recent academic publications. Such a review would offer insights into the historical, recent, and current advancements in recommendation systems. It would deepen our understanding of the influence of filter bubbles and pave the way for new research directions aimed at mitigating their effects on content recommendations. However, the existing literature falls short in terms of in-depth discussions and insightful studies specifically exploring the presence of filter bubbles in RSSs \cite{curkovic2020re}.

This systematic literature review represents the first comprehensive study of its kind that investigates the presence of filter bubbles in RSs. The primary objective of this review is to synthesize and organize the latest research contributions in the field of filter bubbles, employing a well-defined methodology to enhance understanding in this area. The study focuses on classifying existing contributions, evaluating their strengths and weaknesses, and identifying dominant research areas and trends.
Through an extensive review supported by relevant literature and related studies, this review identifies the causes of filter bubble occurrence and examines reported approaches to address this issue. It also proposes potential future research directions to effectively tackle filter bubbles in RSs. Furthermore, it offers a critical assessment of techniques employed to mitigate the negative consequences of filter bubbles, aiming to avoid or reduce their harmful effects. In addition, this paper explores alternative approaches and proposes theoretical models that aim to minimize the influence of filter bubbles on recommendation systems.
The key contributions of this article can be summarized as follows:

\begin{itemize}
\item This study presents the first Systematic Literature Review (SLR) dedicated to investigating the presence of filter bubbles in RSs. It fills a significant gap in the existing research by providing a comprehensive analysis of the literature on this topic.
\item The article examines existing frameworks and provides detailed insights into their features, advantages, disadvantages, and the techniques employed for detecting and mitigating filter bubbles. This analysis helps in understanding the current state of the field and identifying effective strategies for addressing this issue.
\item The article highlights open research issues that need to be addressed to effectively tackle the concerns raised by filter bubbles. These issues provide a roadmap for future investigations and prompt researchers to explore innovative solutions.
\item Additionally, the paper proposes potential research directions that have the potential to contribute significantly to the field in the near future. These directions serve as a valuable resource for researchers looking to expand on the existing knowledge and make further advancements. 
\end{itemize}

{\color{black}\section{Related Works}
\subsection{Recommendation Systems (RSs)}}
Recommender systems (RSs) play a crucial role in providing personalized suggestions to users based on their past interactions. These systems encompass a wide range of recommendations, including movies, products, travel options, advertisements, and news. User preferences can be inferred from their behavior, which can be either implicit or explicit.
Implicit preferences are deduced from activities such as online shopping, website visits, link clicks, and web browser cookies, without directly soliciting feedback from users. On the other hand, explicit feedback involves actively requesting users to provide ratings or comments on the recommendations they have received \cite{ref1}.
Content-based filtering, collaborative filtering, and hybrid approaches are the three most commonly employed recommendation techniques in RSs \cite{ref2,ref3}. Commercial recommendation methods often adopt a combination of these approaches rather than relying solely on content or collaborative filtering. They frequently integrate knowledge-based and context-based strategies to enhance the accuracy and effectiveness of recommendations \cite{ref2}.

The distinction between a current experience and one that has already occurred can be described as novelty, while the internal variations within the components of an experience are referred to as diversity. Initially, recommender systems (RSs) were primarily designed to predict users' interests. However, as research on RSs progressed, the literature began to emphasize a broader perspective on recommendation utility, which includes not only prediction accuracy \cite{ref53,ref54}, but also the importance of originality, variety, and other features in enhancing the value of recommendations \cite{ref55,ref56}. This awareness has grown over time, leading to a surge of activity in this area over the past decade \cite{ref57,ref58,ref59,ref60,ref61}. As a result, novelty and diversity have gained prominence and are increasingly recognized as important evaluation measures for new recommendation systems. Algorithmic advancements are consistently aimed at improving these aspects.

{\color{black}\subsection{Filter Bubble}}
In recent decades, the rise of the Internet has sparked considerable scholarly interest in its potential negative effects on society and the public sphere \cite{ref4}. The concept of the internet filter bubble has gained widespread recognition as a manifestation of this pessimistic perspective. The underlying premise of an echo chamber is that social media users deliberately interact with like-minded individuals and consume content that aligns with their ideologies. As a result, they rarely encounter diverse viewpoints that are crucial for fostering a more inclusive and vibrant public sphere \cite{ref5}.
\begin{table*}[ht]
\centering
\begin{tabular}{|l|l|c|}
\hline
\textbf{Author/Year}                                                            & \textbf{Survey prime coverage}                                                                                                                      & \textbf{\begin{tabular}[c]{@{}c@{}}Related to\\ recommender\\ system?\end{tabular}}  \\ \hline
\cite{ref7}                                                                      & Filter bubble                                                                                                                                       & No                                                                                         \\ \hline
\cite{ref8}                                                        & Selectivity of exposure preferences and actual exposure                                                                                             & No                                                                                         \\ \hline
\cite{ref9}                                                                  & Avoiding filter bubbles in social networks                                                                                                          & No                                                                                         \\ \hline
\cite{ref10}                                                              & Social media echo chambers                                                                                                                          & No                                                                                       \\ \hline
\cite{ref11}                                                          & \begin{tabular}[c]{@{}l@{}}The potential relevance of digital echo chambers \\ and filter bubbles for nature conservation   practice\end{tabular}   & No                                                                                       \\ \hline
\cite{ref12}                                                           & Digital political economy                                                                                                                           & No                                                                                        \\ \hline
\begin{tabular}[c]{@{}l@{}} \cite{ref13} \end{tabular} & \begin{tabular}[c]{@{}l@{}}Risk of echo chambers and filter bubbles on \\ role government institutions, tech companies \\ and scholars\end{tabular} & No                                                                                       \\ \hline
\cite{ref15}                                                                     & Effects of filter bubbles on democracy                                                                                                              & No                                                                                        \\ \hline
\textbf{Our Study}                                                              & \textbf{Filter Bubble in Recommendation System}                                                                                                     & \textbf{Yes}                                                                            \\ \hline
\end{tabular}

\caption{\label{tab:relatedWorks}Brief summary of recent published filter bubble surveys.}
\end{table*}

This phenomenon is exacerbated by the algorithmic content selection employed by social media platforms, which tends to limit users' exposure to novel and diverse content. As a result, online communities become clustered and polarized, lacking the necessary viewpoint diversity.
The concept of the "Filter Bubble" refers to the potential consequence of personalized internet customization, where individuals are isolated from diverse perspectives and information. Users often find themselves exposed to familiar content or consistent information on similar topics, reinforcing their existing knowledge. This concern initially arose in 2009 when platforms like Google began prioritizing customized search results, leading to variations in outcomes for users based on their previous interactions, expressed preferences, and other criteria \cite{ref2}.

Consumers now encounter a more personalized online environment that delivers content tailored to their perceived interests and the preferences of like-minded individuals within their network. While recommendation engines effectively identify users' preferred choices, they can also contribute to information polarization and restrict novelty and variety, exerting a significant influence on user preferences and satisfaction. Consequently, users are exposed to a narrower range of information and content, as recommended and selected options are reinforced, ultimately leading to the formation of information cocoons. In the field of media communication, this phenomenon is commonly referred to as "echo chambers" \cite{ref62}, while information retrieval scholars label it as "filter bubbles." Filter bubbles represent self-reinforcing systems that isolate individuals from diverse ideas, beliefs, or content \cite{ref63}. The filter bubble effect facilitates the solidification of existing beliefs and preferences, potentially leading to the adoption of more extreme views or behaviors over time, a phenomenon known as "group polarization" \cite{ref64}.
In the business context, the filter bubble effect gives rise to the "Matthew effect" among popular items, wherein products and information that deviate from the long tail hypothesis are not recommended, resulting in reduced sales diversity and potential limitations to corporate success \cite{ref65,ref66}. Furthermore, the prevalence of the filter bubble effect in society can lead to the polarization of political ideas and undermine democratic fairness \cite{ref67,ref68}. Additionally, filter bubbles indirectly contribute to the dissemination of undesirable content on online social media platforms, such as rumors and fake news \cite{ref69}. Current recommendation algorithms primarily prioritize enhancing recommendation accuracy rather than promoting diverse outcomes, which is one of the factors contributing to the formation of filter bubbles \cite{ref70}.
While several surveys have been conducted in recent years to explore filter bubbles and recommendation algorithms, no single study comprehensively investigates all the necessary changes required in recommendation systems to address filter bubbles. Most of the research discussed in this section consists of unstructured surveys, and relevant literature pertaining to the review of filter bubbles is also included within this domain.

In 2019, \cite{ref7} presented a critical analysis of the "filter bubble" hypothesis, arguing that its continued emphasis has diverted scholarly attention from more pressing areas of investigation. The authors also highlight the tangible effects of the persistent use of these notions in mainstream media and political discussions, shaping societal institutions, media and communication platforms, and individual users. Traditional broadcast media's diminishing influence in determining information exposure has given way to contemporary information filters such as recommender systems, aggregators, search engines, feed ranking algorithms, bookmarked websites, and the individuals and organizations followed on social media platforms like Twitter. Critics express concerns that the combination of these filters may isolate individuals within their own information bubbles, making it challenging to correct any false beliefs they acquire. In \cite{ref8}, the authors delve into the research surrounding exposure selectivity preferences and actual exposure to shed light on this topic.
Furthermore, \cite{ref9} presents an integrated solution model aimed at assisting users in avoiding filter bubbles within social networks. The author conducted a comprehensive literature review, identifying 571 publications from six highly regarded scientific databases. After removing irrelevant studies and conducting an in-depth analysis of the remaining publications, a recommended category of research papers was developed. This categorization serves as the basis for designing an integrated tool that incorporates previous research findings and introduces novel features to mitigate the impact of filter bubbles.

In 2021, \cite{ref10} conducted a comprehensive review of scientific literature on the subject of echo chambers in social media, aiming to provide a consolidated and critical perspective on the various techniques, similarities, differences, benefits, and limitations associated with echo chambers. This review serves as a foundation for future research in this field. The authors performed a systematic review of 55 studies that examined the presence of echo chambers on social media platforms, classifying the literature and identifying common themes in the focus, techniques, and conclusions of the studies.
Similarly, in their paper, \cite{ref11} provide an exploratory overview of the utilization of digital echo chambers and filter bubbles in the context of nature conservation practice. They gathered data from a literature review and a digital expert poll of German conservation actors to analyze the current understanding of these phenomena. The findings indicate that these concepts are already being investigated in relation to conservation issues, particularly climate protection, and to a lesser extent, natural conservation practice. However, there is a limited understanding of the specific mechanisms underlying digital echo chambers and filter bubbles. The study highlights the urgent need for research and strategic assessment in managing and addressing these challenges in the field of nature conservation.
Furthermore, \cite{ref12} conducted a semi-systematic literature review to examine the digital political economy. They identified and characterized four major threats: false news, filter bubbles/echo chambers, online hate speech, and surveillance. The authors also proposed a typology of "workable solutions" to address these risks, emphasizing the tendency to adopt technological, regulatory, and culturally ingrained approaches as part of the solution.

In \cite{ref13}, the authors conducted a survey of empirical research in the Netherlands to explore tailored information delivery, with a particular focus on echo chambers and filter bubbles in a global context. The study investigated the involvement of government agencies, tech businesses, and academics in addressing these issues. Currently, the Dutch journalism landscape seems to offer a diverse range of information to different citizen groups. However, the precise impact of news personalization is not fully understood, and the increasing influence of digital corporations underscores the need for further research and deeper insights. Without a comprehensive understanding of the situation, it is challenging to develop effective strategies to mitigate the potential concerns of news customization.
Similarly, in \cite{ref14}, a qualitative approach was employed to propose new research directions on the impact of filter bubbles on democracy. The study included a comprehensive literature review and secondary data analysis. The authors argued that the emerging financial models of digital media, heavily reliant on technology companies, marketers, and the public, contribute significantly to the creation of filter bubbles. Newsrooms increasingly gather and analyze customer data for personalized information in digital advertising and subscription models. The media industry enthusiastically embraces customization, with limited critique of its negative aspects. The authors suggested that journalism has a crucial role to play in combating information bubbles by reassessing its digital economic models and raising public awareness.

The previous review of the literature reveals a significant gap in systematic and comprehensive research specifically dedicated to investigating the filter bubble phenomenon in recommendation systems. To fill this gap, we conducted an extensive and critical investigation into the presence and impact of filter bubbles in recommendation systems. This research aims to contribute to the advancement of knowledge and understanding in the fields of recommendation systems and computational social science, offering valuable insights for both researchers and practitioners.
To provide a clear overview of the existing literature reviews in this area, Table \ref{tab:relatedWorks} presents a summary of relevant studies, highlighting their scope, survey methodologies employed, and the number of references considered in each study. This table serves as a reference point for understanding the scope and depth of previous research efforts in this field.

\section{Methodology}
A well-executed survey entails a systematic review and comprehensive analysis of all relevant studies and research conducted on the topic of interest. As highlighted in \cite{ref15}, there are several key motivations for conducting a systematic survey, including synthesizing existing literature and findings on a specific issue, identifying gaps or limitations in the research, and proposing potential avenues for further investigation. By employing a structured and systematic methodology, this type of survey enhances the overall rigor and reliability of the study, allowing for the categorization and analysis of relevant themes and parameters. In this section, we examine the methodological approach employed in our review, highlighting its robustness and its ability to address the objectives and expected outcomes of the study.

\begin{figure*}[ht]
\centering
\includegraphics[width=0.9\linewidth]{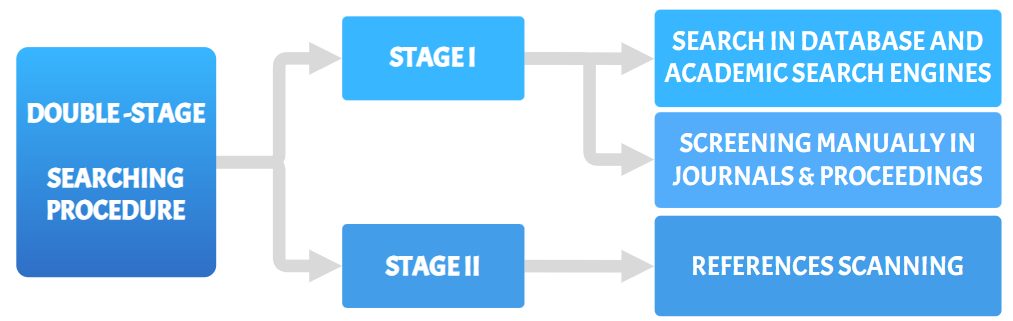}
\caption{Implemented search procedure.}
\label{fig:searchProcess}
\end{figure*}

\subsection{Research questions}
This review article aims to address several key research questions related to the filter bubble phenomenon in recommendation systems. These research questions are as follows:
\begin{enumerate}
\item Does the filter bubble exist in recommendation systems? If so, what are the reasons for its existence?
\item What are the approaches used to identify the presence of a filter bubble in recommendation systems?
\item How can the impact of the filter bubble be mitigated or avoided in recommendation systems?
\end{enumerate}

A systematic literature review (SLR), as described by Kitchenham, is a methodological approach that involves thoroughly examining and synthesizing all relevant works pertaining to a specific research topic or subject area. Systematic reviews provide an objective and comprehensive analysis of the topic by following a rigorous and transparent process that can be audited and replicated. Despite the importance of the filter bubble phenomenon in recommendation systems, a comprehensive systematic literature review specifically focusing on this topic is currently missing in the existing literature. Therefore, conducting a thorough and meticulous analysis, guided by an SLR methodology, is crucial to examine and shed light on the various assertions and findings related to the research questions stated above in an unbiased and replicable manner \cite{ref15}.

\subsection{Bibliographic databases selection criteria}

We conducted an extensive search to gather relevant literature for this systematic literature review (SLR). We focused on recent publications available in reputable scientific journals and top conference proceedings, utilizing leading academic databases. Our search covered the period from 2012 onwards, as our findings indicated that significant research on the filter bubble phenomenon emerged during this time.
To ensure a comprehensive search, we employed a double-staged search strategy consisting of Phase 1 and Phase 2 (see Figure \ref{fig:searchProcess}). In the first phase, we manually explored various search strings and their combinations using Boolean operators. Additionally, we leveraged research databases and academic search engines to access relevant literature from multiple disciplinary and publishing platforms. Figure \ref{fig:selection} provides an overview of the databases, libraries, and search engines that were included in our search strategy.

During the initial phase, we refined our search strings by incorporating specific keywords, including terms from the titles, abstracts, and relevant keywords of publications. This iterative process allowed us to fine-tune our search results and address any potential limitations in matching our searches for thoroughness and consistency. The refined search strings were then applied in the selected databases to retrieve additional relevant literature.

\begin{figure*}[ht]
\centering
\includegraphics[width=\linewidth]{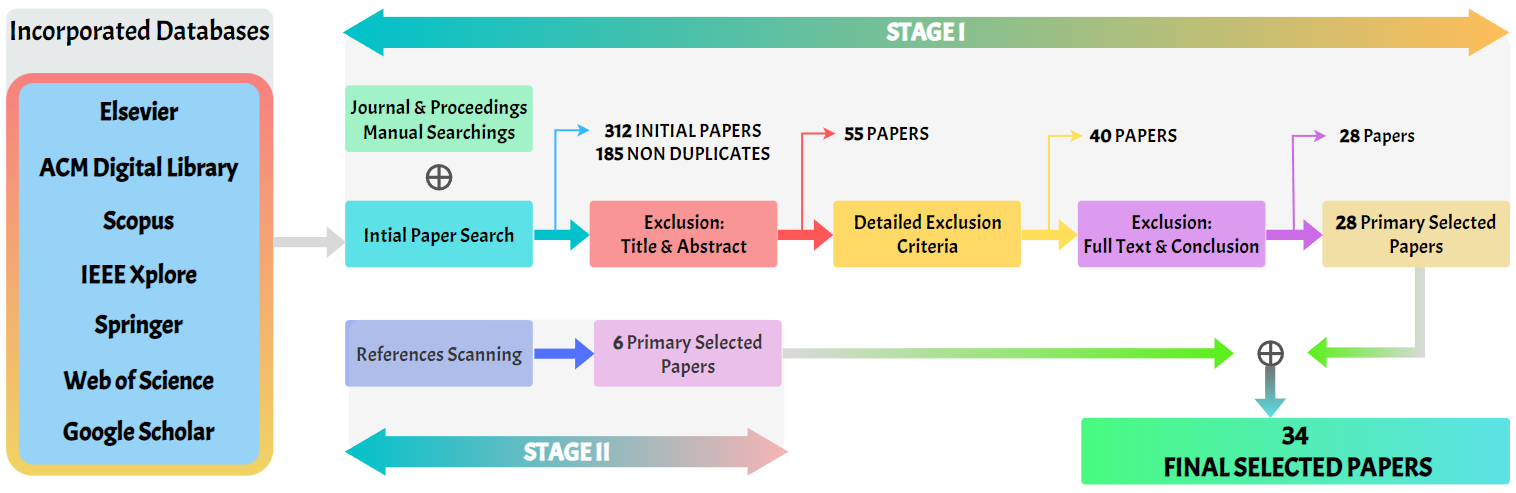}
\caption{Study selection criteria.}
\label{fig:selection}
\end{figure*}

In addition to our systematic search strategy, we conducted manual searches in esteemed journals and established conferences that are highly relevant to our research discipline. These journals and conferences encompass a wide range of topics, including AI, Neural Networks, Recommendation Systems, and related advancements in the relevant disciplines. By including earlier published research from these sources, we aimed to ensure comprehensive coverage of the existing literature and gather valuable insights from the forefront of the field.

\subsection{Search strategy generation}
\textcolor{black}{In this study, we implemented a systematic search strategy to identify pertinent literature addressing filter bubble approaches in recommendation systems. To ensure comprehensive coverage, we established inclusion and exclusion criteria based on expressive and descriptive terms associated with recommender systems and the filter bubble phenomenon. These terms encompassed concepts such as "Recommender System," "Recommendation System," "Filter Bubble," "Echo Chamber," "Self Loop," and other closely related terms. By employing this approach, we aimed to tailor our search to our specific research objectives and review scope, while mitigating the potential impact of nomenclature discrepancies.Subsequently, we employed the Boolean OR operator to consolidate synonyms and related terms as part of our search strategy. This approach aimed to broaden the search scope and encompass various regions where the concept of the filter bubble has been investigated. By using the Boolean OR operator, we aimed to attain comprehensive results while avoiding redundancy. To further refine and narrow down the search outcomes, we then utilized the Boolean AND operator. This step allowed us to focus specifically on studies that concurrently addressed both recommender systems and the filter bubble phenomenon, ensuring the inclusion of relevant literature.}

\subsection{Inclusion and exclusion selection criteria}
\textcolor{black}{During our systematic review, we initially collected 312 papers. To ensure the relevance of the literature, we applied basic criteria such as title, abstract, and topic alignment with our research question. We then established detailed inclusion and exclusion criteria to streamline the selection process. The inclusion criteria encompassed papers proposing solutions, addressing the existence of the filter bubble, implementing techniques, or proposing enhanced versions of recommender systems to mitigate its effects. Conversely, the exclusion criteria were applied to exclude publications that did not specifically address the filter bubble, focused on other applications or research sectors, or compared various recommendation techniques.
One author took the lead in the selection strategy and conducted the initial screening, ensuring consistency with our research theme. Any disagreements regarding the suitability of specific works were resolved through discussions with other authors. After removing duplicates, we identified 185 unique articles. We then conducted a thorough assessment of the remaining articles by carefully reviewing titles, abstracts, and conclusions. Based on this assessment, we narrowed down the selection to 55 articles that exhibited relevance based on title and abstract. In the subsequent stage, we applied the specified inclusion and exclusion criteria to the remaining articles, leading to the exclusion of certain studies that did not meet our criteria. }

\textcolor{black}{
Typically, during the selection process, we applied the following exclusion criteria to refine our literature set:
\begin{itemize}
\item Duplicate records
\item Papers that did not comment on the existence of a filter bubble.
\item Papers related to the implementation of applications utilizing previous RSs.
\item Papers related to research sectors other than RSs.
\item Papers that compared various recommendation techniques.
\item Papers written in languages other than English.
\end{itemize}}
\textcolor{black}{Furthermore, the following inclusion criteria were used to select relevant literature:
\begin{itemize}
\item Proposes a solution to the issue of filter bubbles in recommender systems.
\item Comments on the existence of a filter bubble.
\item Implements a technique or method to alleviate filter bubbles.
\item Proposes an enhanced version of recommender systems to address the problem of filter bubbles.
\end{itemize} }

\textcolor{black}{
By applying these exclusion and inclusion criteria, we ensured that the selected articles provided insights, solutions, or advancements specifically related to the filter bubble phenomenon in RSss. This process resulted in a final selection of 28 articles that met our inclusion criteria. In order to ensure a comprehensive review, we conducted a reference scan of the selected articles, which led us to identify an additional 6 relevant papers. Consequently, a total of 34 articles were included in our systematic review on the existence of the filter bubble. Figure \ref{fig:selection} provides an overview of our research selection criteria and the distribution of publications obtained from each database.}

\begin{figure*}[t!]
\centering
\includegraphics[width=1.0\linewidth]{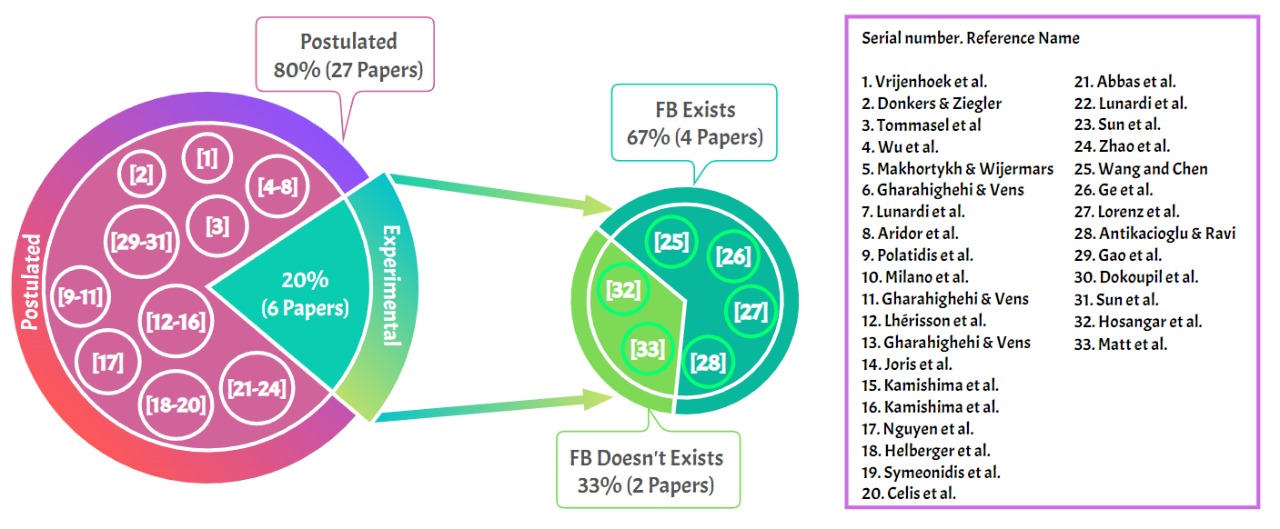}
\caption{Division of studies.}
\label{fig:division}
\end{figure*}

\textcolor{black}{\section{Discussion and Findings}}

\textcolor{black}{Recommender systems (RSs) have the power to either create or dismantle filter bubbles, playing a significant role in shaping the openness or closedness of the internet. However, when analyzing RSs, some methods focus on short-term user engagement and the number of clicks, rather than considering the user's long-term interest in diverse and relevant information. In recent years, researchers have proposed various theories and conducted studies to explore the presence of filter bubbles and echo chambers within RSs. By examining the issues addressed, the techniques employed, and the data used, we can gain insights into the findings and draw conclusions accordingly.
Upon evaluating the data, a clear distinction emerged between studies that identified the presence of filter bubbles and those that did not. We can categorize these studies into three groups: (i) those that found evidence of a filter bubble, (ii) those that did not explicitly comment on its existence, and (iii) those that did not find evidence of a filter bubble but observed heterogeneity, cross-cutting interactions, and exposure. To further analyze the literature, we classified the research based on the methodologies or approaches employed to support their claims. Consequently, we divided the research into two categories: (i) studies that empirically established the presence or absence of the filter bubble, and (ii) studies that assumed its existence or non-existence and utilized it to propose or support another concept.
By examining these categories and the corresponding research, we can gain a deeper understanding of the filter bubble phenomenon and its implications in the context of RSs.}

\textcolor{black}{
When comparing the methodologies, data, and research focuses with the corresponding findings, notable patterns and trends emerged (refer to Table \ref{tab:division} and Figure \ref{fig:division}). Among the collected research, a majority of studies (n = 29) acknowledged the presence of the filter bubble and proposed solutions or alternative theories to address it. Specifically, three out of the 25 experiments provided empirical evidence supporting the existence of the filter bubble. In contrast, only two studies concluded that filter bubbles do not occur. Additionally, five studies did not explicitly comment on the existence of the filter bubble.
These findings highlight the consensus among researchers regarding the prevalence of the filter bubble phenomenon in recommendation systems. The empirical evidence from a subset of experiments further strengthens the argument for its existence. However, it is important to note that the research landscape also includes studies that explore alternative perspectives and propose differing viewpoints. The diversity of approaches and conclusions contributes to a comprehensive understanding of the filter bubble phenomenon and provides insights for future research directions. }

\begin{table*}[ht]
\caption{ Division of studies}
\label{tab:division}
\centering
\small
\begin{tabular}{|l|cccc|}
\hline
\multirow{3}{*}{\textbf{Methods}} & \multicolumn{4}{c|}{\textbf{Findings}}                                                                                                                                                                                                                                         \\ \cline{2-5} 
                                  & \multicolumn{2}{c|}{ \cellcolor{lightgray} \textbf{Filter bubble exists}}                                                                                                                                                                                                                                                                                                                                                          & \multicolumn{2}{c|}{\cellcolor{pink} \textbf{Filter bubble does not exist}}                                                                \\ \cline{2-5} 
                                  & \multicolumn{1}{c|}{\textbf{\begin{tabular}[c]{@{}c@{}}Number of \\ Studies\end{tabular}}} & \multicolumn{1}{c|}{\textbf{Studies}}                                                                                                                                                                                                                                                               & \multicolumn{1}{c|}{\textbf{\begin{tabular}[c]{@{}c@{}}Number of \\ Studies\end{tabular}}} & \textbf{Studies}  \\ \hline
\textbf{Experimentally}           & \multicolumn{1}{c|}{4}                                                                     & \multicolumn{1}{c|}{\cite{ref17,ref30,ref39,ref40}}                                                                                                                                                                                                                                            & \multicolumn{1}{c|}{2}                                                                     & \cite{ref36,ref37}\\ \hline
\textbf{Postulated}               & \multicolumn{1}{c|}{27}                                                                    & \multicolumn{1}{c|}{\begin{tabular}[c]{@{}c@{}} \cite{ref16,ref18,ref19,ref21}\\
\cite{ref22,ref23,ref24,ref25}\\
\cite{ref27,ref28,ref29,ref31} \\
\cite{ref32,ref33,ref34,ref35,ref38}\\
\cite{ref41,ref42,ref43,ref48}\\
\cite{ref49,ref50,ref51,ref52,ref74, ref75}\\
\end{tabular}} & \multicolumn{1}{c|}{0}                                                                     & -                 \\ \hline
\end{tabular}
\end{table*}

Other investigations \cite{ref18, ref19, ref23, ref25, ref27, ref29, ref31, ref32, ref34, ref35, ref41, ref42, ref43, ref48, ref49, ref74, ref75} have also identified the presence of the filter bubble and proposed solutions to address this issue. These studies employed various experimental approaches to devise their solutions.
For instance, \cite{ref23} focused on building diversity-aware neighborhood-based session-based recommender systems. They proposed strategies to diversify the recommendation lists of these systems. The findings revealed that all tested scenarios led to increased diversity across all news databases. The selection of a diversification strategy can be considered as a hyperparameter based on the validation set. Diversification contributes to combating the filter bubble by increasing the number of distinct news topics in the recommendation lists.
Similarly, \cite{ref29} introduced techniques to enhance variety and accuracy in session-based recommender systems using sequential rule mining and session-based k nearest neighbor algorithms. They developed a performance balancing technique to address the filter bubble, which improved the diversity and accuracy of these session-based recommender systems. Real-world datasets from the field of music recommendation were utilized to validate their approach.

Other techniques explored in the literature focused on the usage of the MovieLens dataset, which is a synthetic dataset derived from real-world movie ratings. To address the limitations associated with this dataset, several studies, including \cite{ref27}, \cite{ref34}, and \cite{ref35}, employed experimental techniques.
For instance, Polatidis et al. \cite{ref27} conducted experiments using various recommendation algorithms, ranging from collaborative filtering to complex fuzzy recommendation systems, to tackle the filter bubble problem. They validated their approach using a real-world dataset, and the results indicated its practicality and effectiveness.
Similarly, \cite{ref34} and \cite{ref35} proposed a filter-free recommendation system that promotes information neutrality from a user-defined perspective. They suggested methods to improve the neutrality of the recommendation process, allowing users to have more control over their exposure to diverse content.
In another study, \cite{ref42} utilized multiple MovieLens datasets to propose two models: popularity-based and distance-based Novelty-aware Matrix Factorization (NMF). These models aimed to strike a balance between matrix factorization performance and the need for novelty in recommendations, while only marginally sacrificing accuracy.
Furthermore, \cite{ref31} developed a recommendation model and evaluated it using two publicly available datasets. The results demonstrated that their approach outperformed existing diversification methods in terms of recommendation quality.

In their study, \cite{ref32} propose three scenarios to enhance the diversification of the session-based k-nearest neighbor strategy and address the filter bubble phenomenon. The findings, based on three different news data sources, demonstrate that these diversification scenarios increase the rank and relevance-sensitive diversity metric within the session-based k-nearest neighbor approach.
In order to decrease polarization, \cite{ref43} present a framework that aims to mitigate the formation of echo chambers. Additionally, \cite{ref18} propose a graphical agent-based model to diversify suggestions, promoting exposure to a wider range of information.
Addressing the issue of filter bubbles, \cite{ref41} investigate the construction of recommendations to encourage diverse information exposure and challenge the formation of potential filter bubbles.

In the context of social media, \cite{ref19} suggest an echo chamber-aware buddy recommendation algorithm based on Twitter data. This algorithm learns individual and echo chamber representations from shared content and previous interactions of users and communities.
Examining the recommendation environment, \cite{ref25} explore situations where consumers remain within their filter bubbles despite receiving diverse recommendations. They find that while recommendations can mitigate the effects of filter bubbles, they may also lead to user boredom, resulting in a trade-off between diversifying across users and within-user consumption.
In the domain of diet diversification, \cite{ref48} develop a case-based reasoning (CBR) system called DiversityBite. This system promotes diet diversification by generating dynamic criticism that guides users through different search areas and encourages them to explore alternative examples. The authors evaluate the impact of DiversityBite on diversity through user research in the recipe domain.

\cite{ref49} addressed the filter bubble issue, specifically focusing on the role of recommender systems in causing it within the News domain. To tackle this challenge, they developed a point-of-view diversification technique. This technique stands out as the first functional and active News recommender system that incorporates point-of-view diversity, distinguishing it from previous studies.
Similarly, \cite{ref50} proposed an adaptive diversity regularization CDMF (Collaborative Deep Matrix Factorization) model. Their approach utilizes social tags as a means to connect the target and source domains, resulting in improved recommendation accuracy and enhanced recommendation diversity through adaptive diversity regularization. To evaluate the effectiveness of their proposed methodology, extensive experiments were conducted on a real social media website. The analysis of the data led to several important conclusions. Firstly, the use of social tags to overcome the low recommendation accuracy caused by the target domain's sparsity proved to be particularly beneficial. Secondly, the incorporation of adaptive regularization significantly increased the individual variety of recommendations. Lastly, their proposed methodology struck a fair balance between accuracy and diversity of recommendations, while also reducing user polarization.

\textcolor{black}{Only two studies included in this analysis reported no evidence of a filter bubble in recommendation systems. These studies found that recommendation systems actually help users broaden their interests and create commonalities with other users. Both studies employed different approaches to analyze personalization and focused on its positive aspects.
For instance, \cite{ref36} examined data from an online music service and found that personalization does not lead to fragmentation of the online population. Instead, they observed that as users follow recommendations, their purchasing behavior becomes more similar to that of other users, as indicated by purchase similarity.
Similarly, \cite{ref37} found that perceived suggestion serendipity has a significant positive impact on both perceived preference fit and user satisfaction. Their findings suggest that simply increasing the number of innovative recommendations is not enough. Instead, recommenders should make occasional random suggestions, which can lead to a higher perception of preference fit and enjoyment for users.}

\textcolor{black}{\subsection{Existence of filter bubble}}
In this section, we present the overall results of our study, which are based on the persuasive research, observed trends, comparative analysis, and analytical assessment conducted by all authors through a thorough debate and deliberation. Based on our findings, we have observed that research in the field of the filter bubble is growing. While the number of studies on the filter bubble is still relatively small due to its emerging nature, there has been a significant increase in research activity in recent years. As depicted in Figure \ref{fig:distribution}, which illustrates the annual distribution of filter bubble studies, there were only 8 publications from 2012 to 2018, whereas in 2021 alone, there were 9 publications on the topic.
Through various methodologies and datasets, the presence of a filter bubble in recommendation systems has been convincingly demonstrated. The studies have examined contextual biases using diverse datasets and platforms. Furthermore, the majority of investigations successfully illustrated the personalized effect of recommendation systems. Therefore, based on the literature we reviewed, we can confidently conclude that the filter bubble exists in recommendation systems.

\begin{figure*}[ht]
\centering
\includegraphics[width=0.8\linewidth]{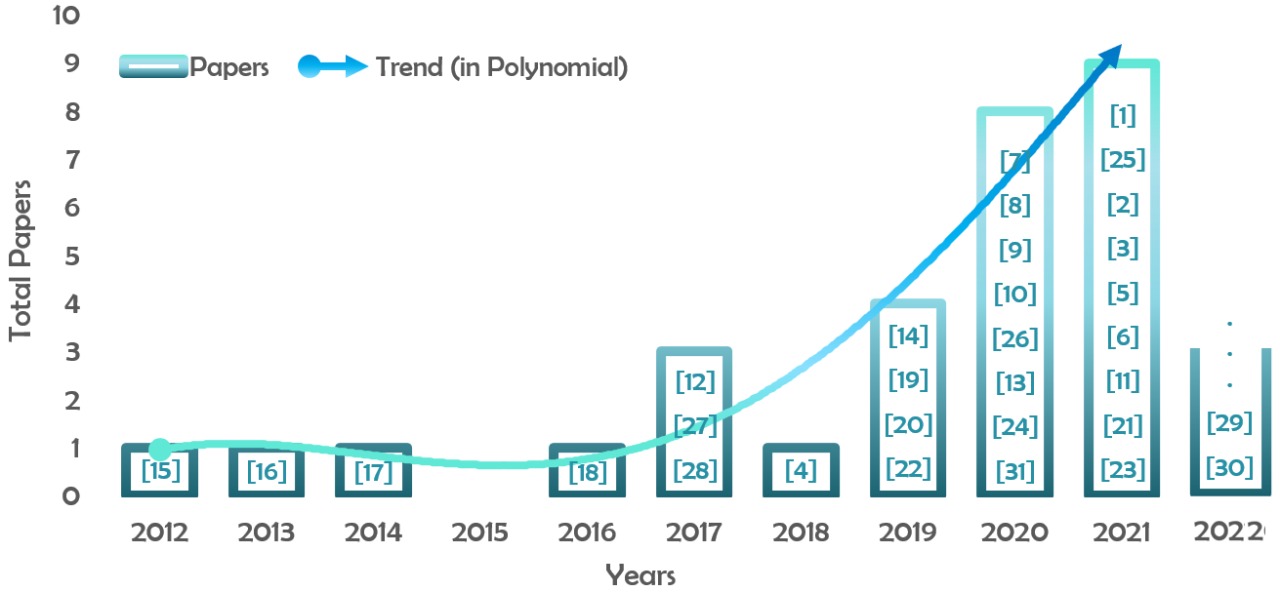}
\caption{Yearwise distribution of all studies}
\label{fig:distribution}
\end{figure*}

The literature extensively examines various forms of bias that contribute to the problem of personalization in recommendation systems (RSs). Biases can arise at different stages, including during system design and implementation, evaluation, and user interaction. These biases can significantly impact the information gathered for system improvement and customization \cite{ref45}.
One prominent form of bias is algorithmic bias, which refers to biases introduced during the design and implementation of the RS. This bias can be a result of the underlying algorithms and data processing techniques used in the system. Additionally, biases can arise from the evaluation process, where researchers may unknowingly introduce their own biases into the assessment of the system's performance. The design of the user interaction is also critical, as it can introduce additional biases in the form of presentation or exposure bias \cite{ref45}.
Furthermore, cognitive biases, such as confirmation bias and other behavioral biases, can influence the user's interactions with the system and introduce biases into the data collected. These biases can affect the feedback loops used by RSs, as they are based on implicit user feedback, such as clicks or other trackable user activities. However, due to the limitations of these feedback mechanisms, the interactions are skewed towards the options presented by the system, leading to a form of bias known as presentation or exposure bias \cite{ref45}.
According to the research, the major causes of filter bubbles in recommendation systems can be attributed to algorithmic bias, data bias, and cognitive bias. These biases can have significant implications for the personalization and customization of RSs, and addressing them is crucial to mitigate the formation of filter bubbles.

\subsection{Approaches to identifying FB}

 \par {\color{black}SSeveral research studies in the literature have proposed strategies to understand, avoid, and mitigate the harmful effects of the filter bubble phenomenon (refer to Table \ref{tab:experimental} and \ref{tab:postulated}). This category of research explores novel ideas and diverse perspectives on how to identify and counteract the negative impact of recommendation algorithms that contribute to the formation of filter bubbles.
Different approaches have been employed to determine the existence of a filter bubble, with studies utilizing benchmark datasets such as MovieLens, Twitter, or self-generated datasets. For instance, \cite{ref39} conducted their research using a user interaction dataset from a WebTV platform and demonstrated that contextual bias leads to biased program recommendations, resulting in users being trapped in a filter bubble. To address this, they leveraged the Twitter social stream as an external context source, expanding the selection to include content related to social media events. They investigated the Twitter histories of key programs using two trend indicators: Trend Momentum and SigniScore. The analysis showed that Trend Momentum outperformed SigniScore, accurately predicting 96 percent of all peaks in the selected candidate program titles ahead of time.
While many studies rely on datasets to support their research, some propose frameworks or models without utilizing specific datasets. For example, \cite{ref43} proposed a generic framework to prevent polarization by ensuring that each user is presented with a balanced selection of content. They demonstrated how modifying a basic bandit algorithm can improve the regret bound above the state-of-the-art while satisfying the requirements for reducing polarization.
These research studies offer valuable insights and methodologies for understanding and addressing the filter bubble phenomenon, providing a foundation for developing effective strategies to mitigate its negative effects in recommendation systems.}

\begin{table*}[ht]
\centering
\footnotesize
\begin{tabular}{|c|c|l|c|c|}
\hline
\textbf{Refs} & \textbf{Year} & \textbf{Dataset Used}                                   & \textbf{\begin{tabular}[c]{@{}c@{}}Approach to\\ identify\end{tabular}} & \textbf{\begin{tabular}[c]{@{}c@{}}Solution \\ proposed?\end{tabular}} \\ \hline
\cite{ref17}      & 2021          & Large survey data collected from e-commercial platforms & Algorithmic                                                                   & No                                                                     \\ \hline
\cite{ref30}      & 2020          & Alibaba Taobao                                          & Statistical                                                                   & No                                                                     \\ \hline
\cite{ref39}      & 2017          & Twitter                                                 & Trend Detection                                                               & Yes                                                                    \\ \hline
\cite{ref40}      & 2017          & MovieLens-1m, and Netflix   Prize data                  & Graphical                                                                     & Yes                                                                    \\ \hline
\end{tabular}
\caption{\label{tab:experimental} Analysis of experimental approaches}
\end{table*}

\begin{table*}[]
\caption{ Analysis of postulated approaches.}
\label{tab:postulated}
\centering
\scriptsize
\begin{tabular}{|c|c|l|l|c|c|}
\hline
\textbf{Refs} & \textbf{Year} & \textbf{Dataset Used}                                                                                                               & \textbf{What did authors propose?}                                                            & \textbf{Approach}                                                     & \textbf{\begin{tabular}[c]{@{}c@{}}Solution \\ proposed ?\end{tabular}} \\ \hline
\cite{ref16}     & 2021          & -                                                                                                                                   & Set of metrics                                                                                & Theoretical                                                           & No                                                                      \\ \hline
\cite{ref18}     & 2021          & -                                                                                                                                   & Agent-based model                                                                             & Graphical                                                             & Yes                                                                     \\ \hline
\cite{ref19}      & 2021          & Twitter                                                                                                                             & \begin{tabular}[c]{@{}l@{}}Echo chamber-aware friend\\  recommendation System\end{tabular}    & Modeling                                                              & Yes                                                                     \\ \hline
\cite{ref21}      & 2018          & Douban Interest Group   dataset                                                                                                     & \begin{tabular}[c]{@{}l@{}}Personality-based greedy\\ re-ranking approach\end{tabular}        & Experimental                                                          & No                                                                      \\ \hline
\cite{ref22}      & 2021          & Manual Data Collection                                                                                                              & Theories about filter bubble                                                                  & Analysis                                                              & No                                                                      \\ \hline
\cite{ref23}      & 2021          & \begin{tabular}[c]{@{}l@{}}Roularta   Kwestie Globo.com\\ Adressa\end{tabular}                                                      & \begin{tabular}[c]{@{}l@{}}Scenarios to diversify the \\ recommendation lists\end{tabular}    & Experimental                                                          & Yes                                                                     \\ \hline
\cite{ref24}      & 2020          & \begin{tabular}[c]{@{}l@{}}Brazilian   presidential elections of\\ 2018 Data\end{tabular}                                           & Metric to measure filter bubble                                                               & Algorithmic                                                           & No                                                                      \\ \hline
\cite{ref25}      & 2020          & Manual Data Collection                                                                                                              & Model                                                                                         & \begin{tabular}[c]{@{}c@{}}Numerical\\    \\ simulation\end{tabular}  & Yes                                                                     \\ \hline
\cite{ref27}      & 2020          & MovieLens 1 million                                                                                                                 & Explanation-based approach                                                                    & Experimental                                                          & Yes                                                                     \\ \hline
\cite{ref28}      & 2020          & -                                                                                                                                   & \begin{tabular}[c]{@{}l@{}}Analysis of social effects of\\  filter bubble\end{tabular}        & Analysis                                                              & No                                                                      \\ \hline
\cite{ref29}      & 2021          & \begin{tabular}[c]{@{}l@{}}Real-life datasets from the music\\ recommendation domain\end{tabular}                                   & Performance balancing approach                                                                & \begin{tabular}[c]{@{}c@{}}Empirical\\    \\ evaluations\end{tabular} & Yes                                                                     \\ \hline
\cite{ref31}      & 2017          & MovieLens and Last.fm dataset                                                                                                       & Build a recommendation model                                                                  & Modeling                                                              & Yes                                                                     \\ \hline
\cite{ref32}      & 2020          & Roularta1, Globo.com and Adressa                                                                                                    & News RS                                                                       & Algorithmic                                                           & Yes                                                                     \\ \hline
\cite{ref33}      & 2019          & -                                                                                                                                   & News RS                                                                       & Algorithmic                                                           & No                                                                      \\ \hline
\cite{ref34}      & 2012          & Movielens 100k                                                                                                                      & RS                                                                            & Experimental                                                          & Yes                                                                     \\ \hline
\cite{ref35}      & 2013          & Movielens 100k                                                                                                                      & RS                                                                            & Experimental                                                          & Yes                                                                     \\ \hline
\cite{ref38}      & 2014          & MovieLens                                                                                                                           & \begin{tabular}[c]{@{}l@{}}Metric to measure content \\ diversity\end{tabular}                & Analysis                                                              & No                                                                      \\ \hline
\cite{ref41}      & 2016          & -                                                                                                                                   & \begin{tabular}[c]{@{}l@{}}Three normative conceptions \\ of  exposure diversity\end{tabular} & Analysis                                                              & Yes                                                                     \\ \hline
\cite{ref42}      & 2019          & \begin{tabular}[c]{@{}l@{}}MovieLens 100K (ML100K), \\ MovieLens 1ML (ML1M), \\ MovieLens 20 ML (ML20ML) and\\  Yelp 6\end{tabular} & Two Models                                                                                    & Experimental                                                          & Yes                                                                     \\ \hline
\cite{ref43}      & 2019          & Curated dataset of online news articles                                                                                             & Framework                                                                                     & Algorithmic                                                           & Yes                                                                     \\ \hline
\cite{ref48}      & 2021          & Recipe Dataset                                                                                                                      & RS                                                                            & Experimental                                                          & Yes                                                                     \\ \hline
\cite{ref49}     & 2019          & -                                                                                                                                   & Representation model                                                                          & Experimental                                                          & Yes                                                                     \\ \hline
\cite{ref50}      & 2021          & Douban Dataset                                                                                                                      & Model                                                                                         & Experimental                                                          & Yes                                                                     \\ \hline
\cite{ref51}      & 2020          & -                                                                                                                                   & Agent-Based simulation                                                                        & Framework                                                             & Yes                                                                     \\ \hline
\cite{ref74}      & 2022          & Reddit and Yelp                                                                                                                                   & RS                                                                       & Experimental                                                             & Yes                                                                     \\ \hline
\cite{ref75}      & 2022          & -                                                                                                                                  & RS                                                                       & Modeling                                                             & Yes                                                                     \\ \hline
\end{tabular}

\end{table*}

\par {\color{black}Examining users' behavior is another important aspect of identifying the filter bubble phenomenon. For instance, \cite{ref39} incorporated the Twitter social stream as an external context source to expand the selection of items to include those related to social media events. They recognized the significance of users' behavior in determining the composition of the filter bubble.
Similarly, \cite{ref17} investigated the biases of four algorithms based on five metrics (relevance, variety, novelty, unexpectedness, and serendipity) across user groups categorized by eight different characteristics. To gain insight into the identified biases, they analyzed users' behavioral patterns, such as their inclination to provide more favorable ratings. The study found that biases varied to a greater extent among user groups based on their age and curiosity levels.
Despite the range of research projects conducted in this area, there is a common observation that real-time implementation of the proposed methodologies in recommendation systems has received limited attention. The practical application and integration of these research findings into real-world recommendation systems have been identified as an important area for future exploration and development.}

\par {\color{black}Graph/network-based analysis and visualization have been employed by researchers to investigate the presence of the filter bubble. For instance, \cite{ref19} developed FRediECH, a system that combines echo chamber awareness with user representations to balance the relevance, diversity, and originality of friend suggestions. FRediECH utilizes a Deep Wide architecture and a graph convolutional network to enhance the diversity of recommendations by re-ranking the results based on the network's explicit community structure. However, this approach may have limitations as it requires defining the criteria for identifying such groups. FRediECH aims to adapt the community structure to changes in user interactions and content patterns, striking a balance between relevance and variety.
In another study, \cite{ref23} employed a CNN-based deep neural network technique to construct article embeddings for news articles using information such as article title, synopsis, full text, and tags from datasets. They utilized the Maximal Marginal Relevance (MMR) re-ranking technique, which compares the results of the suggested approaches with a diversified baseline. The MMR-based method evaluates multiple performance criteria, such as accuracy and variety, to re-rank items from the original recommendation list. While MMR-based methods help reduce the impact of the filter bubble, they are often criticized for being computationally expensive and sacrificing relevance for diversity, making them less feasible in real-world scenarios.
Addressing these concerns, \cite{gao2022mitigating} proposed a novel approach called Targeted Diversification VAE-based Collaborative Filtering (TD-VAE-CF) to mitigate political polarization in media recommendations. This approach aims to strike a balance between relevance and diversity by leveraging the capabilities of Variational Autoencoders (VAE) in generating diverse and targeted recommendations.
}

\begin{table*}[h]
\caption{Various approaches to solve filter bubble in the literature.}
\label{tab:solution}
\centering
\small
\begin{tabular}{|c|ccc|l|}
\hline
\multirow{3}{*}{\textbf{Ref}} & \multicolumn{3}{c|}{\textbf{Solution Approach}}                                                                                                            & \multirow{3}{*}{\textbf{Technique used}}                                                                 \\ \cline{2-4}
                              & \multicolumn{1}{c|}{\textbf{Re-ranking}} & \multicolumn{1}{c|}{\textbf{\begin{tabular}[c]{@{}c@{}}Diversity\\    Modeling\end{tabular}}} & \textbf{Other} &                                                                                                          \\ \hline
\cite{ref18}                      & \multicolumn{1}{c|}{\text{\ding{55}}}                   & \multicolumn{1}{c|}{\text{\ding{51}}}                                                                         & \text{\ding{55}}              & Knowledge Graph Embedding                                                                                \\ \hline
\cite{ref19}                      & \multicolumn{1}{c|}{\text{\ding{51}}}                   & \multicolumn{1}{c|}{\text{\ding{55}}}                                                                         & \text{\ding{55}}              & Graph Convolutional Networks                                                                             \\ \hline
\cite{ref23}                      & \multicolumn{1}{c|}{\text{\ding{51}}}                   & \multicolumn{1}{c|}{\text{\ding{55}}}                                                                         & \text{\ding{55}}              & Convolutional Neural   Network                                                                           \\ \hline
\cite{ref25}                      & \multicolumn{1}{c|}{\text{\ding{55}}}                   & \multicolumn{1}{c|}{\text{\ding{51}}}                                                                         & \text{\ding{55}}              & Expected Utility Theory                                                                                  \\ \hline
\cite{ref27}                     & \multicolumn{1}{c|}{\text{\ding{55}}}                   & \multicolumn{1}{c|}{\text{\ding{55}}}                                                                         & \text{\ding{51}}              & Explanations                                                                                             \\ \hline
\cite{ref29}                      & \multicolumn{1}{c|}{\text{\ding{51}}}                   & \multicolumn{1}{c|}{\text{\ding{55}}}                                                                         & \text{\ding{55}}              & \begin{tabular}[c]{@{}l@{}}Sequential   rule mining and \\ session-based k nearest neighbor\end{tabular} \\ \hline
\cite{ref31}                     & \multicolumn{1}{c|}{\text{\ding{51}}}                   & \multicolumn{1}{c|}{\text{\ding{55}}}                                                                         & \text{\ding{55}}              & Mexican-Hat Diversity Model                                                                              \\ \hline
\cite{ref32}                      & \multicolumn{1}{c|}{\text{\ding{51}}}                   & \multicolumn{1}{c|}{\text{\ding{55}}}                                                                         & \text{\ding{55}}              & Convolutional Neural   Network                                                                           \\ \hline
\cite{ref34}                      & \multicolumn{1}{c|}{\text{\ding{55}}}                   & \multicolumn{1}{c|}{\text{\ding{51}}}                                                                         & \text{\ding{55}}              & Latent Factor Model                                                                                      \\ \hline
\cite{ref35}                      & \multicolumn{1}{c|}{\text{\ding{55}}}                   & \multicolumn{1}{c|}{\text{\ding{51}}}                                                                         & \text{\ding{55}}              & Probabilistic Matrix Factorization Model                                                                 \\ \hline
\cite{ref39}                      & \multicolumn{1}{c|}{\text{\ding{55}}}                   & \multicolumn{1}{c|}{\text{\ding{55}}}                                                                         & \text{\ding{51}}              & External Social Context                                                                                  \\ \hline
\cite{ref40}                      & \multicolumn{1}{c|}{\text{\ding{51}}}                   & \multicolumn{1}{c|}{\text{\ding{55}}}                                                                         & \text{\ding{55}}              & Graphical                                                                                                \\ \hline
\cite{ref41}                      & \multicolumn{1}{c|}{\text{\ding{55}}}                   & \multicolumn{1}{c|}{\text{\ding{55}}}                                                                         & \text{\ding{51}}              & Suggestions                                                                                              \\ \hline
\cite{ref42}                      & \multicolumn{1}{c|}{\text{\ding{55}}}                   & \multicolumn{1}{c|}{\text{\ding{51}}}                                                                         & \text{\ding{55}}              & Matrix Factorization                                                                                     \\ \hline
\cite{ref43}                      & \multicolumn{1}{c|}{\text{\ding{55}}}                   & \multicolumn{1}{c|}{\text{\ding{55}}}                                                                         & \text{\ding{51}}              & Simple Bandit Algorithm                                                                                  \\ \hline
\cite{ref48}                      & \multicolumn{1}{c|}{\text{\ding{55}}}                   & \multicolumn{1}{c|}{\text{\ding{51}}}                                                                         & \text{\ding{55}}              & Critique-Based Conversational Recommendation                                                             \\ \hline
\cite{ref49}                      & \multicolumn{1}{c|}{\text{\ding{55}}}                   & \multicolumn{1}{c|}{\text{\ding{51}}}                                                                         & \text{\ding{55}}              & Natural Language Processing                                                                              \\ \hline
\cite{ref50}                      & \multicolumn{1}{c|}{\text{\ding{55}}}                   & \multicolumn{1}{c|}{\text{\ding{51}}}                                                                         & \text{\ding{55}}              & Adaptive Diversity   Regularization                                                                      \\ \hline
\cite{ref51}                      & \multicolumn{1}{c|}{\text{\ding{55}}}                   & \multicolumn{1}{c|}{\text{\ding{51}}}                                                                         & \text{\ding{55}}              & Agent-Based Simulation                                                                                   \\ \hline
\cite{ref52}                      & \multicolumn{1}{c|}{\text{\ding{55}}}                   & \multicolumn{1}{c|}{\text{\ding{51}}}                                                                         & \text{\ding{55}}              & Adaptive Diversity   Regularization                                                                      \\ \hline
\cite{ref74}                      & \multicolumn{1}{c|}{\text{\ding{55}}}                   & \multicolumn{1}{c|}{\text{\ding{51}}}                                                                         & \text{\ding{55}}              & Trains Concept Activation Vectors                                                                     \\ \hline
\cite{ref75}                      & \multicolumn{1}{c|}{\text{\ding{55}}}                   & \multicolumn{1}{c|}{\text{\ding{51}}}                                                                         & \text{\ding{55}}              & Long-Term fairness                                                                     \\ \hline
\end{tabular}
\end{table*}

\par {\color{black} After identifying the presence of filter bubbles, many studies have proposed potential solutions. The first category of solutions focuses on bypassing or modifying algorithms. In our selected research, a significant number of solutions concentrated on enhancing content diversity. For instance, \cite{ref23} and \cite{ref32} presented scenarios to make session-based recommendation systems more diversity-aware by considering not only a user's current session interactions but also diverse content from other sessions. Additionally, \cite{ref43} proposed a flexible framework that allows users to have control over the source from which recommendations are selected, thereby reducing polarization in personalized systems. Furthermore, some researchers have identified strategies to enable users to explore fresh information that was previously unknown to them (\cite{ref19}).
To achieve content diversity, the two most commonly used approaches in recommendation systems are re-ranking and diversity modeling. Re-ranking methods, such as those proposed by \cite{ref19}, \cite{ref23}, and \cite{ref29}, involve post-processing techniques that reorder the ranked list provided by the baseline recommender. They assess the diversity of suggestions on the candidate list and perform a re-ranking based on this criterion. While these strategies can enhance diversity, they often require additional post-processing steps and can be computationally expensive. On the other hand, diversity modeling approaches, as suggested by \cite{ref18}, \cite{ref25}, and \cite{ref34}, involve modifying the core algorithm itself to make it more diversity-aware. These approaches adapt the recommendation algorithm to incorporate diversity as a key consideration (see Table \ref{tab:solution}).}

\par {\color{black} Several researchers have explored the incorporation of diversity regularization into matrix factorization (MF) models to achieve multi-objective recommendations that maximize both accuracy and variety. In their study, \cite{ref35} utilize a probabilistic matrix factorization approach (\cite{ref72}) to predict ratings, which has shown significant success in terms of prediction accuracy and scalability. Similarly, \cite{ref42} propose two models, namely popularity-based and distance-based novelty-aware MF, which allow for a trade-off between matrix factorization performance and the requirement for novelty while only moderately sacrificing accuracy. The results of their experiments suggest that it is possible to achieve high accuracy while also introducing unique and diverse recommendations.}

\begin{figure}[ht]
\centering
\includegraphics[width=0.7\linewidth]{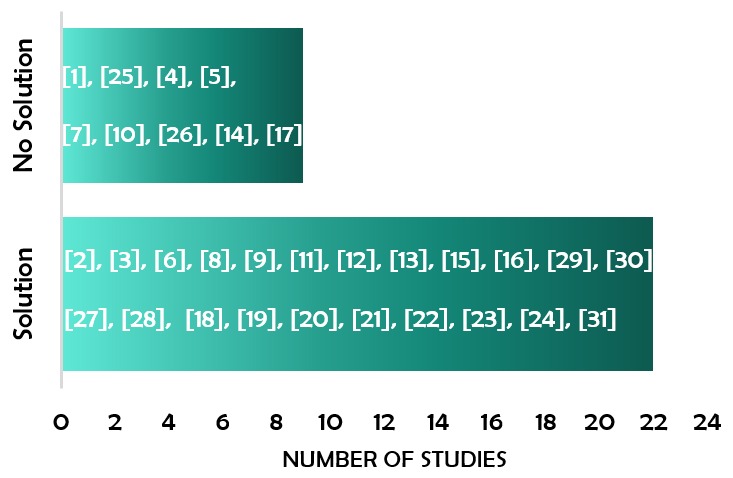}
\caption{Distribution of studies on solution basis}
\label{fig:studyDistribution}
\end{figure}

\par {\color{black} In summary, the majority of research in this area focuses on enhancing diversity in recommendations while still maintaining a level of personalization. Additionally, there is a strong emphasis on making the recommendation process more transparent and explainable, as well as involving users in the decision-making process. Many researchers have also highlighted the importance of developing frameworks or models that are efficient and feasible for real-world scenarios. Building upon these insights, the authors of this study propose generalized methods to mitigate the filter bubble phenomenon in recommender systems, which will be discussed in the next section.}

\textcolor{black}{\section{Preventing filter bubble}}
Despite being a relatively nascent area of research, this study has successfully identified commonalities and variations in the understanding of echo chambers in recommender systems. It provides a comprehensive and critical analysis of peer-reviewed literature, shedding light on this significant issue. The field itself is complex and fragmented, characterized by challenges in collecting, interpreting, and comprehending variables and data. Nevertheless, the importance and potential of studying echo chambers in recommender systems are evident. In the subsequent sections, we will present several viable approaches to addressing the filter bubble problem. We strongly believe that user awareness is a crucial initial step towards mitigating this issue. Informed users can question why certain recommendations are suggested and understand the user features influencing those recommendations. This awareness also empowers users to recognize bias in the presented information and encourages them to explore opposing opinions and recommendations. Additionally, we will propose strategies to tackle the creation of filter bubbles in recommender systems.

\subsection{Modeling filter bubble as multi-objective optimization problem}
We know that the filter bubble is created due to highly personalized recommendations. A possible way to avoid this situation is to add some diversity to the recommendations through various means, including random recommendations. However, we can not completely neglect the personalized recommendations generated through previous user experiences. The solution lies in a balance between personalized and diversified recommendations. Both components are necessary but of competing nature, i.e., increasing one will decrease the other. Such conflict situations can be seen and modeled as a multi-objective optimization problem. The solution to a multi-objective optimization problem is a set of ‘non-inferior’ or ‘non-dominated’ solutions called a Pareto-optimal front.

Theoretically, this set contains infinitely many points for which no solution can be said better than the others. For example, a possible solution Pareto set for filter bubble could be: {100\% personalization, 0\% diversification}, {90\% personalization, 10\% diversification},\dots, {50\% personalization, 50\% diversification},\dots, {0\% personalization, 100\% diversification}. The first solution of the solution set {100\% personalization, 0\% diversification} focuses only on personalized recommendations. On the other hand, the last solution {0\% personalization, 100\% diversification} prefers diverse recommendations only. However, there are many intermediate solutions that try to make a balance between both. An important point to note here is that one solution is not better than any other solution because each has a better value for exactly one objective. The concept of the Pareto optimal set is described in Figure \ref{fig:paretoFront}.

The filter bubble is a result of highly personalized recommendations. To avoid this situation, it is necessary to introduce diversity into the recommendations, which can be achieved through various means, including random recommendations. However, personalized recommendations based on previous user experiences cannot be completely disregarded. The solution lies in finding a balance between personalized and diversified recommendations, recognizing that both components are necessary but inherently compete with each other. This conflict can be formulated and modeled as a multi-objective optimization problem.

In a multi-objective optimization problem, the solution space consists of a set of 'non-inferior' or 'non-dominated' solutions known as the Pareto-optimal front. Theoretically, this set comprises infinitely many points, with no solution being considered better than others. For instance, in the context of addressing the filter bubble, the Pareto set may include solutions such as {100\% personalization, 0\% diversification}, {90\% personalization, 10\% diversification}, \dots, {50\% personalization, 50\% diversification}, \dots, {0\% personalization, 100\% diversification}. The first solution in the set, {100\% personalization, 0\% diversification}, focuses solely on personalized recommendations, while the last solution, {0\% personalization, 100\% diversification}, prioritizes diverse recommendations. However, there exist many intermediate solutions that aim to strike a balance between both objectives. It is important to note that no single solution is superior to others since each solution offers better values for a specific objective. The concept of the Pareto-optimal set is illustrated in Figure \ref{fig:paretoFront}.

\begin{figure}[ht]
\centering
\includegraphics[width=1\linewidth]{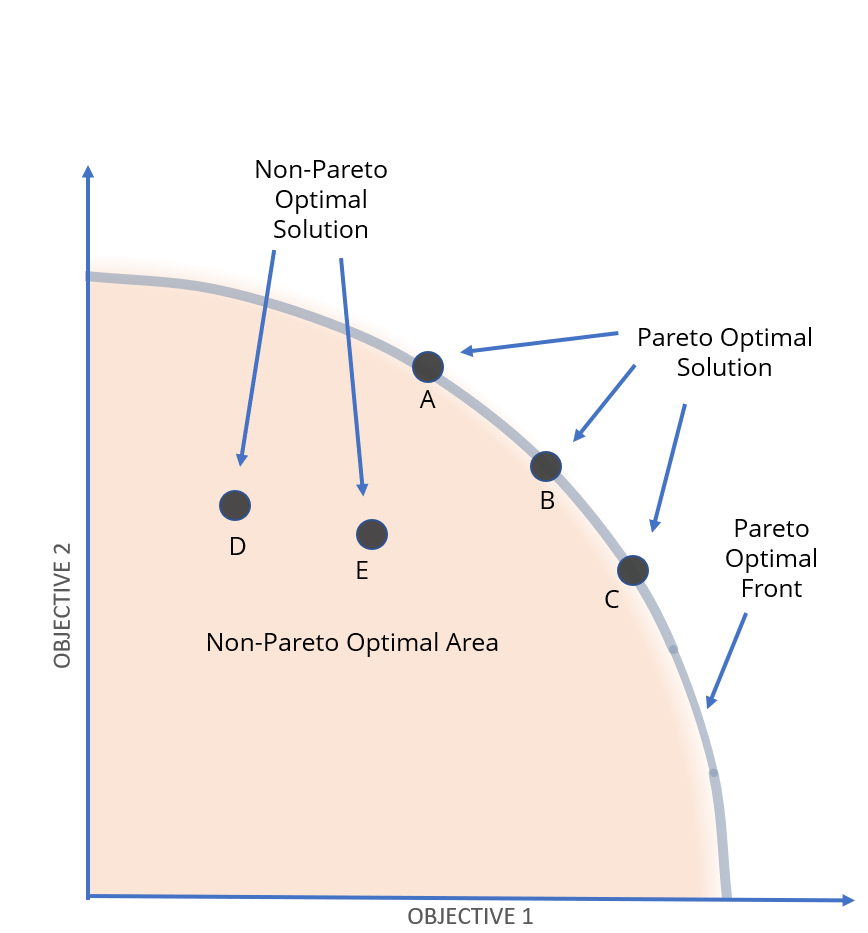}
\caption{Pareto optimal set for a bi-objective maximization optimization problem}
\label{fig:paretoFront}
\end{figure}

Here, the solutions A, B, and C are incomparable but all of them are better than solution D and E. If the filter bubble problem is posed as a bi-objective optimization problem, it may be represented as Eq. \ref{eq:optimize}:

\begin{align}\label{eq:optimize}
\textbf{Maximize}~~ \textit{Diversity Score}\\
\textbf{Maximize}~~ \textit{Personalization Score}\nonumber
\end{align}

The Diversity Score measures the degree of diversified recommendations, while the Personalization Score represents the degree of personalized recommendations in the final outcome, both normalized to the range [0,1]. The Pareto set of Eq. \ref{eq:optimize} is depicted in Figure \ref{fig:fbPareto}.
In this figure, point P (0,1) represents a solution that emphasizes full personalization, while point D (1,0) represents a completely random recommendation. Recommendations A, B, and C fall within the \textit{Desirable Area} of the Pareto-optimal front, exhibiting non-zero values for both scores, but with varying degrees. Recommendation A contains more personalized information than B and C, while C has a higher level of diversity.
Once we have developed such a theoretical model, the next step is to define the mathematical formulation of Eq. \ref{eq:optimize}, which involves determining the formulas for calculating the Diversity Score and the Personalization Score. By solving Eq. \ref{eq:optimize}, we can obtain a set of recommendations that have incomparable values of personalization and diversity scores. Recommendations falling within the \textit{Desirable Area} are expected to generate bubble-free results.

\begin{figure}[ht]
\centering
\includegraphics[width=1\linewidth]{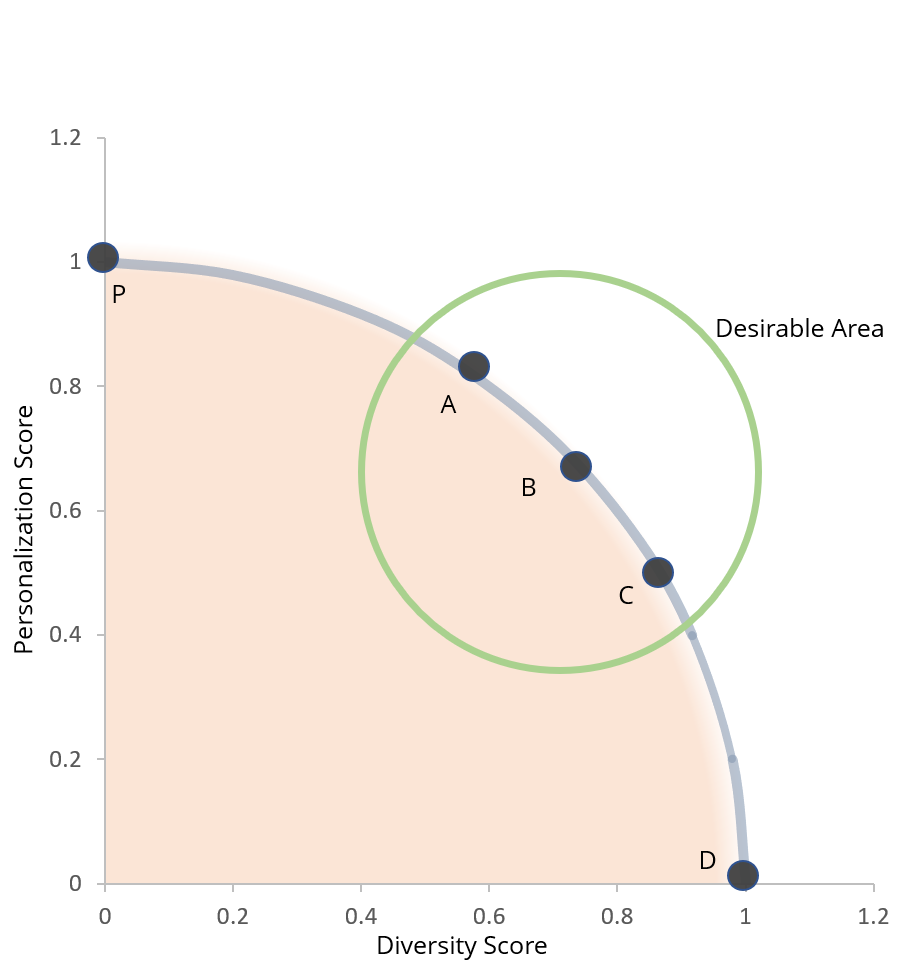}
\caption{Pareto optimal front of filter bubble problem}
\label{fig:fbPareto}
\end{figure}

\subsection{Explainable Recommender Systems (XRSs)}
Based on the insights gained from our research, we propose an architecture for integrated tools that can be employed in recommendation systems to mitigate the formation of filter bubbles. Drawing upon the findings of our literature analysis, we suggest that this integrated tool should serve two primary functions: (1) alerting users to the potential presence of a filter bubble, and (2) allowing users to customize the extent of personalization.

\begin{sidewaysfigure*}
\centering
\includegraphics[width=\linewidth]{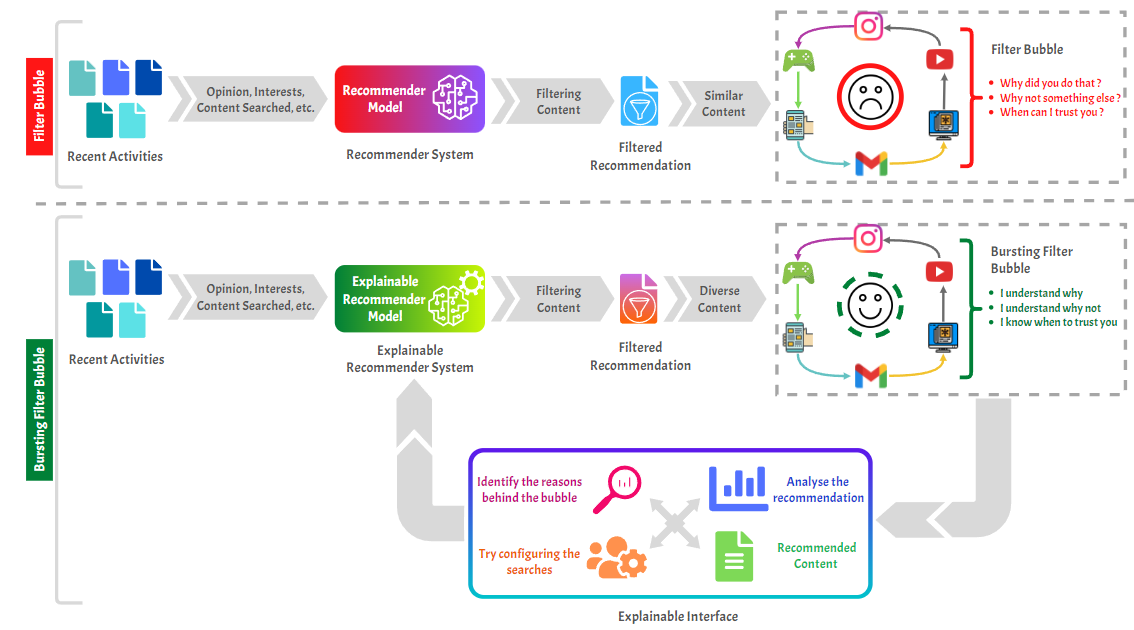}
\caption{An illustration of the effect of XRSs over filter bubble}
\label{fig:tool}
\end{sidewaysfigure*}

\begin{figure}[t!]
\centering
\includegraphics[width=\linewidth]{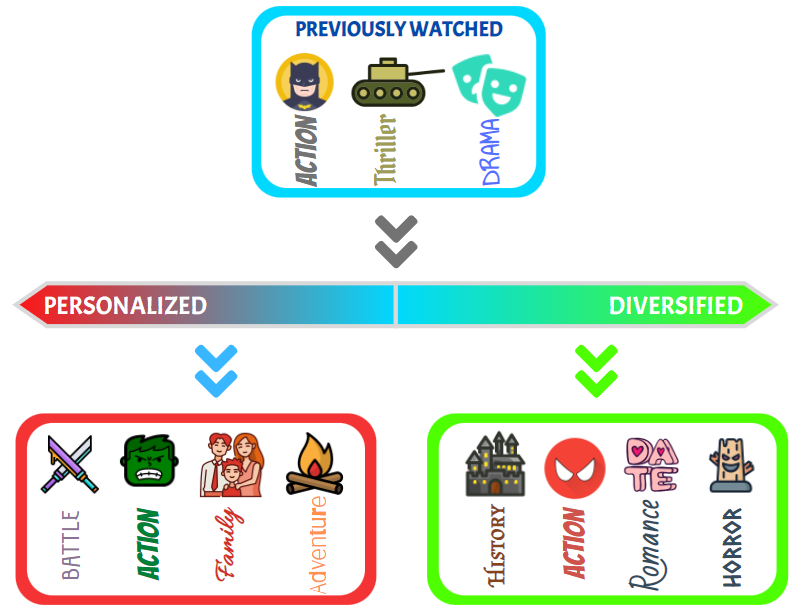}
\caption{An example of personalization and diversification of recommendations}
\label{fig:scale}
\end{figure}

{\color{black}In recent times, there has been a growing interest in explainable artificial intelligence (XAI) across various research domains, aiming to address the challenges posed by increasing complexity, scalability, and automation \cite{varlamis2022smart,ref46}. Consequently, the development of explainable recommendation systems (XRSs) has gained momentum. Notably, researchers such as Peake et al. \cite{peake2018explanation} have proposed a novel approach for extracting explanations from latent factor recommendation systems by employing training association rules on the outcomes of a matrix factorization black-box model. Their method effectively balances interpretability and accuracy without compromising flexibility or relying on external data sources.}
Explanations play a crucial role in ensuring that users comprehend and trust recommendation systems that prioritize explainability. Without accompanying explanations, there is a risk that the recommendations generated by a system may be perceived as untrustworthy or lacking authenticity \cite{ref47}. By understanding the rationale behind a recommendation, users can identify potential filter bubbles and take steps to burst them. For instance, if an item is accompanied by a rating indicating the level of personalization in the suggestion, whether it is based on previous searches or purely random \cite{ashraf2023private}, users can gain insights into why the recommendation is being made.

In line with designing a fair and explainable system, an XRS focused on food recipe recommendations has been proposed \cite{yera2022exploring}. The notable contribution of this recommendation approach is its comprehensive inclusion of explainability features, which not only provide explanations for recommendations but also raise nutrition awareness. By incorporating additional aspects into the explanation process, this approach aims to enhance user satisfaction and understanding, making it a valuable component of an XRS.}

Balancing the trade-off between personalization and diversification is crucial when recommending items in order to address the filter bubble phenomenon. Customized recommendations are important as they facilitate the user's search for relevant items. However, it is equally important to provide diverse results to break the bubble effect. Therefore, we aim to incorporate this trade-off into our tools and give users the ability to choose the type of recommendations they desire.
By providing users with control over this trade-off, the recommendation system can achieve its goal while also preventing users from being trapped in a filter bubble. For instance, if a user prefers items that are similar to their previous searches, the degree of personalization can be adjusted to provide more tailored recommendations. On the other hand, if a user wants to explore a wider range of items without being influenced by their past data, they can modify the degree of personalization to receive more diverse recommendations.

The tool proposed in Figure \ref{fig:tool} aims to provide users with a better understanding of the recommendations they receive and empower them to customize their future searches to break free from the filter bubble. By offering transparency and explanation, users can gain insights into why a specific recommendation was made, allowing them to make informed decisions and challenge the bubble effect.
Figure \ref{fig:tool} also illustrates a comparison between the proposed explainable recommendation system and a standard recommendation system. The added layer of explainability in the proposed system enhances user understanding and trust, promoting a more satisfying user experience.
Figure \ref{fig:scale} focuses on the tool's interface, using a movie suggestion example. In this scenario, the user's preferences primarily revolve around action, thriller, and drama movies, as depicted in the figure. When the system is personalized, the user is presented with recommendations that align with their preferred genres. On the other hand, when the degree of personalization is adjusted towards diversity, the system recommends a broader range of content, allowing the user to explore movies beyond their usual preferences.

\subsection{Approaches for diversification}
As discussed in previous sections, the primary solution to combat the filter bubble problem is to incorporate diverse content in recommendations. However, it is crucial to define diversity itself as it encompasses various types, each with its specific definition and implications. It is worth noting that current recommendation systems intentionally introduce some level of variety to ensure that the recommended items are not excessively similar \cite{smyth2001similarity}. Additionally, other types of diversity, such as personalized and temporal diversity, are also being utilized in recommendation systems \cite{kunaver2017diversity}.
While measures of diversity are already employed in recommendation systems, their objective has not always been to address the filter bubble issue but rather to provide users with a range of somewhat dissimilar options to choose from. Consequently, it becomes crucial to define diversity in the context of the filter bubble phenomenon. In selecting an appropriate diversity measure, several key considerations should be taken into account.

\begin{itemize}
\item \textbf{Opposite of similarity:} In early recommendation systems, diversity was viewed as the opposite of similarity and defined as (1 - similarity), where similarity is a measure of the proximity between user interests and recommended items \cite{smyth2001similarity}. In a list of items, diversity is calculated as the average dissimilarity between all pairs of items.
\item \textbf{Diversity through Rearrangement/Re-ranking:} This approach involves rearranging the list of recommended items generated by the algorithm to improve the diversity metric \cite{adomavicius2011improving}. It has been observed that this simple approach works well in certain scenarios. It can be seen as an optimization problem that aims to maximize the diversity metric.
\item \textbf{Diversity in items and/or source:} It is important to decide whether diversity should be introduced only in the recommended content or in the content provider as well \cite{vrijenhoek2021recommenders}. For example, in online shopping, diversified items may include different garments, while diversified sources may involve different brands.
\item \textbf{Personalized/User-specific Diversity:} Diversity can be introduced irrespective of user profiles, which is referred to as non-personalized diversity. However, it is considered better to capture the diversity needs of individuals by modeling their characteristics and incorporating them into the diversity metric \cite{eftimov2021framework}. Such diversity measures are known as personalized matrices.
\item \textbf{Temporal Diversity:} In certain domains, recommendations need to consider the dimension of time, giving rise to the concept of temporal diversity \cite{xiang2010temporal}. News recommendation systems, for instance, must account for rapidly changing news topics, as well as the evolving preferences of users over different time periods (weekly, monthly, yearly, or seasonally). Thus, temporal diversity should be designed to address users' short- and long-term preferences.
\item \textbf{Hybrid Diversity:} A diversity metric may incorporate multiple aspects discussed above, resulting in a hybrid diversity measure \cite{schafer2002meta}. A simple implementation could involve calculating a weighted sum of various diversity measures to capture different dimensions of diversity.
\end{itemize}

Overall, the process of selecting the right diversity metric is a meticulous task that involves careful consideration of various factors. To ensure an effective selection, the following steps need to be followed:
\begin{enumerate}
\item Study the specific domain of the recommendation system under consideration. This involves understanding the characteristics of the items, the preferences of the users, and any temporal or contextual factors that may influence recommendations.
\item Define diversity in the context of the predetermined domain. This entails identifying the specific dimensions or aspects of diversity that are relevant and meaningful for the given domain.
\item Select appropriate diversity measure(s) that align with the defined notion of diversity. This may involve choosing from existing diversity metrics or developing new ones tailored to the specific requirements of the domain.
\item Combine the selected diversity measure(s) with an appropriate prevention approach to effectively address the filter bubble problem. This could involve incorporating diversity constraints into recommendation algorithms or utilizing post-processing techniques for re-ranking recommendations.
\item Gather feedback from users, either implicitly through user interactions or explicitly through surveys or interviews, to evaluate the effectiveness of the diversity measures and their impact on user satisfaction.
\item Adapt and modify the diversity measure(s) based on the received feedback. This iterative process ensures that the diversity metric continues to capture the evolving needs and preferences of the users.
\end{enumerate}

\section{Open Issues And Future Research Directions}
Several open challenges related to overcoming filter bubble in RSs exist, including but not limited to:
\textcolor{black}{
\subsection{Open issues}
\begin{itemize} 
\item \textbf{Defining diversity in a domain-specific context:} Diversity plays a critical role in addressing the filter bubble problem, but its definition may vary depending on the recommendation domain \cite{marras2022equality}. For instance, diversity in a movie recommendation system may differ from diversity in an online clothing portal. It is important to establish domain-specific definitions of diversity and develop mathematical frameworks accordingly. It is worth noting that similar concepts to diversity, such as novelty and serendipity, have been discussed in the literature \cite{sharma2022explanation, castells2022novelty}. While diversity refers to the presence of variety in a recommended item list, novelty captures the difference between past and present recommendations, and serendipity occurs when new and relevant but previously unknown items are included in the recommendations. The choice of which concept or combination to utilize should be based on the specific requirements of the application.
\item\textbf{Exploring contrasting recommendations or opinions:} When addressing the filter bubble issue, incorporating contrasting recommendations or opinions can promote a more balanced understanding, particularly in news recommendation systems. However, it is necessary to define the concept of "Opposite Recommendations" and establish domain-specific definitions to effectively incorporate this approach. It should be noted that defining "Opposite" is relatively straightforward in domains like news recommendation but may pose challenges in other domains, such as book recommendations \cite{liu2023app}.
\item \textbf{Identifying sources responsible for spreading fake news:} Identifying sources responsible for spreading fake news is crucial in addressing the filter bubble problem. Fake news or misinformation greatly contributes to the issue. However, developing advanced natural language processing (NLP) techniques that can effectively detect fake news poses a challenge, especially when dealing with aspects such as sarcasm and deceptive language usage. Deep learning-based NLP models like Deep Bidirectional Transformers, along with techniques like transfer learning and fine-tuning, can be explored to enhance language understanding and mitigate the negative impact of the filter bubble \cite{ref73}.
\item \textbf{Establishing the relationship between domain-specific external factors and the filter bubble:} Establishing the relationship between domain-specific external factors and the filter bubble is crucial in understanding and addressing this phenomenon. Various external factors, such as the presence of fake news in news recommendation systems, contribute to the filter bubble problem. It is important to investigate and comprehend the connection between these factors and the filter bubble.
Tracing the origins of misinformation is a vital step in addressing the filter bubble, and advanced natural language processing (NLP) techniques can greatly assist in this process. Furthermore, the impact of the filter bubble can vary significantly across different applications. For example, in a food recommendation system, a filter bubble can have detrimental effects on users' well-being by excluding nutritious diets and promoting a particular genre of food. The findings of \cite{yera2022exploring} also highlight how the filter bubble effect can introduce intentional biases when providing choices for restaurants and related domains.
On the other hand, the influence of the filter bubble may be more pronounced in a video streaming platform like YouTube, while having only a marginal effect on users in a dress/outfit recommendation system for an online clothing portal. It is essential to recognize that generalized solutions may not be effective for every application, emphasizing the need for domain-specific analysis of the filter bubble. Each application requires a tailored approach and a deeper understanding of its specific dynamics to effectively mitigate the filter bubble's impact.
\item \textbf{Enhancing data quality for visualization and integration:} Enhancing the quality of data is of utmost importance for effective visualization and integration in the context of the filter bubble. As emphasized by \cite{sardianos2021emergence}, researchers should dedicate efforts to explore methods that can enhance the quality of data used in this context. By improving data quality, we can ensure more reliable and accurate results in visualization and integration processes.
Furthermore, it is crucial to address the issue of information cocooning that is prevalent in news recommender systems. These systems often filter out content that users may find uninteresting, resulting in a narrowing of their information exposure over a period of approximately seven days. This can have significant implications, particularly for individuals who are heavily reliant on social media platforms. It is imperative for the research community to tackle the challenge of designing evaluation mechanisms that incorporate social filtering. By doing so, we can mitigate the potential negative consequences of information cocooning and promote a more diverse and balanced information environment for users \cite{sayed2021intelligent}.
\end{itemize} }

\textcolor{black}{
\subsection{Future Research Directions}
Several promising research directions could be pursued to mitigate the filter bubble problem:
\begin{itemize}
\item \textbf{Diversity-aware recommendations:} Designing algorithms that aim to increase the diversity of recommendations can help in mitigating the filter bubble. These algorithms need to balance the trade-off between relevance and diversity \cite{sardianos2020real,hirata2023categorical}.
\item \textbf{Serendipity in recommendations:} Developing recommendation techniques that emphasize serendipity (unexpected but useful recommendations) could help users discover new, out-of-bubble content. These methods would encourage exposure to diverse and novel items that the user might not have found otherwise \cite{wang2023correction}.
\item \textbf{Explainability and transparency:} Explainable AI can help users understand why a particular recommendation was made. Seeing the rationale behind the recommendations might make users more receptive to different content, reducing the filter bubble effect \cite{atalla2022recommendation}.
\item \textbf{User-controlled recommendations:} Allowing users to have more control over their recommendations, such as adjusting the degree of novelty or diversity, could also help alleviate the filter bubble problem.
\item \textbf{Cross-domain recommendations:} Leveraging data from different domains can help in providing a broader range of recommendations. For example, if a user interacts with various content types (books, movies, music), these can be used to cross-pollinate recommendations across these domains \cite{wang2023correction}.
\item \textbf{Fairness and bias mitigation:} Actively researching and implementing algorithms that take into account and mitigate biases in recommender systems can help to ensure that the system does not favour certain types of content, hence reducing the risk of a filter bubble \cite{shi2023selection,liu2023mitigating}.
\item \textbf{Long-term user modeling:} Traditionally, recommender systems have focused on immediate rewards (clicks, purchases, etc.), leading to a filter bubble. Research into long-term user modeling can help understand the evolving needs and tastes of users, potentially aiding in delivering a more diverse set of recommendations \cite{xu2023optimizing}.
\end{itemize}}

\section{Conclusion}
The term "Filter Bubble" refers to the phenomenon where internet personalization isolates individuals by presenting them with content and perspectives that align with their existing preferences. Consequently, users are exposed to a limited range of information or similar content on related topics. This issue gained attention in 2009 when platforms like Google started customizing search results based on users' previous interactions, expressed preferences, and various other factors \cite{ref4}. Many individuals rely on recommendation systems (RSs) to assist them in finding products that align with their specific needs. While RSs offer numerous benefits, they also have the potential to trap users within a filter bubble due to their heavy reliance on similarity measurements. In this Systematic Literature Review, we investigate the existence, causes, and potential solutions to the filter bubble problem in recommendation systems.

We addressed the research problems by conducting an extensive analysis of the studies reported in the literature. The findings confirm the presence of a filter bubble in recommendation systems. This raises the question: What are the underlying causes of excessive personalization in RSs? The literature points to algorithmic bias and cognitive bias as the primary culprits. Algorithmic bias arises when biases are introduced during the design and implementation of a system, while cognitive biases, such as confirmation bias, taint the interaction data. To address this issue, diversification techniques are commonly employed. In recommendation systems, re-ranking and diversity modeling are the two most prevalent methods of diversification. Re-ranking involves post-processing the ranked list provided by a baseline recommender, but this approach increases the computational complexity of the overall algorithm (\cite{ref19,ref23,ref29}). On the other hand, diversity modeling techniques modify the core algorithm to incorporate diversity-awareness (\cite{ref18,ref25,ref34}).

Our work has made significant contributions in reviewing the existing literature across various domains within recommender systems. We have examined the causes of the filter bubble phenomenon, identified trends, and proposed strategies for its identification and prevention. Our key findings highlight the importance of diversity in recommendations while maintaining personalized experiences, as well as the need for transparency and explainability in the recommendation process.
While recent studies have expanded our understanding of the filter bubble, it is important to note that the complexity of these models often hinders their practical adoption. Taking this into consideration, we have outlined generalized methods that can effectively mitigate the filter bubble issue in recommender systems. One promising approach involves employing multi-objective optimization techniques to strike a balance between personalization and diversification.
In addition, we emphasize the significance of incorporating an explanatory framework that provides users with insights into why a particular item is recommended. To this end, we present the components of an integrated tool in the form of an architectural map, which can aid in the prevention of filter bubbles and enhance user understanding and control over recommendations.

The present study sheds light on several promising research avenues that lie ahead. One important aspect is the establishment of criteria for selecting appropriate definitions of personalization and diversification, along with the development of corresponding mathematical metrics. It is evident that these definitions should take into account the specific characteristics of the domain or application under consideration. In fact, each component of the strategy aimed at mitigating the filter bubble issue should be tailored to the particular domain.
Hence, there is a pressing need to devise domain-specific strategies for resolving the filter bubble problem. Such strategies should address various concerns, including assessing the degree of filter bubble present in the application, understanding its impact, determining the necessity for reduction, identifying suitable measures of personalization and diversification, and selecting appropriate prevention methodologies. By taking a domain-centric approach, we can develop effective solutions that are tailored to the unique challenges and requirements of each application.



\section*{Acknowledgments (not compulsory)}
We would like to acknowledge the support of the UK Engineering and Physical Sciences Research Council (EPSRC) Grants Ref. EP/M026981/1, EP/T021063/1, EP/T024917/.

\section*{Author contributions statement}

The corresponding author (S.S.S) initiated the idea of the review, discussed it with all co-authors, they all contributed in writing and structuring the article. S.S.S, Q.M.A and R.I worked on literature collection via searching over academic databases. The diagrams were suggested by S.S.S and created by R.I., Y.H and A.A supervised the idea and structure of the paper. All authors reviewed and revised the manuscripts. S.S.S, M.N and Q.M.A have contributed equally to the manuscript.

\section*{Competing interests}

The authors declare no competing interests.


\end{document}